# Threshold amplitudes for transition to turbulence in a pipe*

Hua-Shu Dou[1,2], Boo Cheong Khoo[2], and Khoon Seng Yeo[2]

[1]Temasek Laboratories, National University of Singapore, Singapore 119260
[2]Fluid Mechanics Division, Department of Mechanical Engineering
National University of Singapore, Singapore 119260, SINGAPORE
Email: tsldh@nus.edu.sg; huashudou@yahoo.com

**Abstract:** Threshold amplitude of disturbance for transition to turbulence in a pipe Poiseuille flow is investigated. Based on the energy gradient theory, we argued that the transition to turbulence depends on magnitudes of the energy gradient of mean flow and the disturbance energy. Furthermore, the threshold disturbance energy required for turbulence transition is expressed as a function of the complementary angle to the energy angle which also characterizes the behaviour of the energy gradient of mean flow. With some mathematical treatments, the variation of the threshold energy of disturbance versus the Reynolds number is obtained. Then, it is found for a fixed disturbance frequency that the normalized amplitude of disturbance is scaled by Re with an exponent of -3/2 for the transition occurrence in a pipe. This value of exponent agrees well with several recent results reported in the literature. Finally, the mechanism of transition to turbulence is suggested for different disturbance levels.

___________________________________________________________________________

**Contents**

___________________________________________________________________________

## 1. Introduction

Turbulence is a very complex phenomenon which has a long history of more than 120 years. The purported mechanism of turbulence generation is very diverse, and the understanding of the flow physics of turbulence is still limited despite the ensuing years of works[1,2]. The pipe Poiseuille flow has been studied by many researchers since Reynolds' famous experiment showed flow transition from laminar to turbulence [3-6]. It is generally accepted from experiments that there is a critical Reynolds number $\text{Re}_c$ below which little or no turbulence can be produced or sustained regardless of the level of imposed disturbance. For pipe Poiseuille flow, this critical value of Reynolds number is about 2000 from experiments [3,4]. Above this $\text{Re}_c$, the transition to turbulence depends to a large extent on the initial disturbance to the flow. For example, experiments showed that if the disturbances in a laminar flow can be carefully avoided or considerably reduced, the onset of turbulence is delayed to Reynolds numbers up to Re=O($10^5$) [7,8]. Experiments also showed that for Re>$\text{Re}_c$, only a threshold of disturbance is reached can transition to turbulence be effected [6]. It has been suggested that the critical disturbance amplitude leading to transition varies broadly with the Re by an exponent rule, i.e., $A \propto \text{Re}^\gamma$. The magnitude of this exponent has significant importance for turbulence research [7,8].

Trefethen et al [7] suggested that the exponent $\gamma < -1$ for plane Couette and pipe Poiseuille flows. Lundbladh et al [9]'s direct numerical simulations showed disturbances whose amplitudes scale with a exponent of -5/4 and -7/4 for the plane Couette and Poiseuille flows, respectively. Baggett et al [10] proposed a linear model for the transition, and their numerical results showed a scaling with an exponent of -3; however, they still surmised that the exponent behaves as $\gamma < -1$. Waleffe [11], based on his discussion on available analytical and numerical data, claimed that for Couette flow the exponent is near -1. On the other hand, Baggett and Trefethen [12] concluded that the threshold exponent for "transition" is strictly less than -1. Chapman [13] through a formal asymptotic analysis of the Navier-Stokes equations found that the exponent is -1 and -3/2 for plane Couette flow and plane Poiseuille flow, respectively. In more recent time, from Darbyshire and Mullin [6]'s experiments, Trefethen et al [14] proposed that the amplitude should be normalized by the average velocity U, and obtained $\gamma = -3/2$ for pipe flows. Messeguer [15] carried out simulations and the results obtained seem to support the scaling of $\gamma = -3/2$ which compared reasonably with the experiments of Darbyshire and Mullin [15]. Recently, Hof et al [16] using pulse disturbance in experiments, obtained the normalized disturbance flow rate in the pipe as inversely proportional to the Re number, i.e., $\gamma = -1$.

However, experiments also showed that the threshold amplitude ($A$) is related to the frequency of the disturbance [6]. For a given Re, the higher the $\omega$, the smaller is the $A$. This result demonstrated that the critical condition is not solely determined by the disturbance amplitude, but by both the amplitude and the frequency. This suggests that the magnitude of the disturbance energy may yet be a key factor affecting transition because frequency is related to the energy. The detailed behaviour and dynamics of the disturbance energy action is very complex.

Currently, all of the major theories of flow instability, including linear theory, energy method, weak nonlinear theory, and second instability theory, can not fully or satisfactorily explain the critical Reynolds number problem of flow instability in parallel flows [1, 2]. Recently, Dou [17] proposed a new theory for flow instability and turbulence transition based on energy gradient concept. This theory uses an energy gradient parameter as a criterion for the subcritical transition and the said parameter suggests a consistent value for plane Poiseuille flow, pipe Poiseuille flow and plane Couette flow for the critical condition of turbulent transition. The energy gradient parameter can be also expressed by the so-called "energy angle," see Fig.1. Since the critical value of the energy gradient parameter in the theory can only be observed from experiments, this theory is a semi-empirical theory.

In this paper, based on the energy gradient theory, we explore the transition condition in the high Re range. We argued that the transition to turbulence depends on the energy gradient of mean flow and the disturbance energy. Furthermore, it is assumed that the energy of disturbance is proportional to the complementary angle of the energy angle, and the variation of critical amplitude versus the Reynolds number is derived. As such, the effect of frequency of disturbance is shown since the disturbance energy is proportional to the square of the disturbance frequency. The theory is then compared with experiments and the results from other researchers.

**2. Energy gradient theory**

Linear stability theory represents quite the state-of-the-art of research for flow stability problems; it obtains agreement with experiments for some problems such as Rayleigh-Benard and Taylor-Couette flows and partial agreement with experiment for plane boundary layer problem. However, the subcritical condition of turbulent transition predicted by linear stability theory differs largely from the experiments for wall bounded parallel flows. Other theories including energy method, weak nonlinear theory, and second instability theory, also can not obtain satisfactory agreement with experiments either for this class of wall bounded flows [1,2,7,8].

Partly motivated by the above stated problem, Dou [17] proposed an *energy gradient theory* with the aim to clarify the mechanism of transition from laminar to turbulence for wall bounded shear flows. Here, we give a short discussion for a better understanding of the work presented in this study. In the theory, the whole flow field is treated as an energy field. It is thought that the gradient of total energy in the transverse direction of the main flow and the viscous friction in the streamwise direction dominate the instability phenomena and hence the flow transition for a given disturbance. The energy gradient in the transverse direction has the potential to amplify a velocity disturbance, while the viscous friction loss in the streamwise direction can resist and absorb this disturbance. The transition to turbulence depends on the relative magnitude of these two roles of energy gradient amplification and viscous friction damping to the initial disturbance. Based on such, a new dimensionless parameter, $K$ (the ratio of the energy gradient in the transverse direction to that in the streamwise direction), is defined to characterize the stability of the base flow,

$$K = \frac{\partial E / \partial n}{\partial E / \partial s}. \tag{1a}$$

Here, $E = p + \frac{1}{2}\rho V^2 + \rho g \xi$ is the total energy for incompressible flows with $\xi$ as the coordinate perpendicular to the ground, $n$ denotes the direction normal to the streamwise direction and $s$ denotes the streamwise direction. $\rho$ is the fluid density, g the gravity acceleration, V the velocity, and p the hydrodynamic pressure. As is well known, this energy of fluid (E) is derived from Bernoulli's equation. For pressure driven flows, the magnitude of energy gradient in streamwise direction equals to the rate of energy loss of unit volume fluid along the streamline due to the viscous friction, $\partial E / \partial s = \partial H / \partial s$. In other word, the mechanism of generation of streamwise energy gradient is resulted by the energy loss due to viscous friction. For shear driven flows, the calculation of K can be obtained by the ratio of the energy gradient in transverse direction and the rate of energy loss of unit volumetric of fluid along the streamline,

$$K = \frac{\partial E / \partial n}{\partial H / \partial s}. \tag{1b}$$

Here, $H$ is the energy loss per *unit volumetric fluid* along the streamline for finite length which has same dimension as $E$. As such, the parameter $K$ in Eq.(1) is a field variable. Thus, the distribution of $K$ in the flow field and the property of disturbance may be the perfect means to describe the disturbance amplification or decay in the flow. It is suggested that the flow instability can first occur at the position of $K_{max}$ which is construed to be the most "dangerous" position. Thus, for a given disturbance, the occurrence of instability depends on the magnitude of this dimensionless parameter $K$ and the critical condition is determined by the maximum value of $K$ in the flow. For a given flow geometry and fluid properties, when the maximum of $K$ in the flow field is larger than a critical value $K_c$, it is expected that instability can occur for certain initial disturbance [17]. The analysis has suggested that the transition to turbulence is due to the energy gradient and the disturbance amplification, rather than just the linear eigenvalue instability type as in stated in [7, 8], because the pipe Poiseuille flow is linearly stable for all the Reynolds number. Both Grossmann [8] and Trefethen et al. [7] commented that the nature of the onset-of-turbulence mechanism in parallel shear flows must be different from an eigenvalue instability of linear equations of small disturbance. In fact, finite disturbance is needed for the turbulence initiation in the range of finite Re as found in experiments [6]. Dou [17] demonstrated that the criterion obtained has a consistent value at the subcritical condition of the transition determined by the experimental data for plane Poiseuille flow, pipe Poiseuille flow as well as plane Couette flow (see Table 1). From this table it can be deduced that the turbulence transition takes place at a consistent critical value of $K_c$ at about 385 for both the plane Poiseuille flow and pipe Poiseuille flow, and about 370 for plane Couette flow. This may suggest that the subcritical transition in parallel flows takes place at a value of $K_c \approx 370\text{-}385$. The finding further suggests that the flow instability is likely resulted from the action of energy gradients, and not due strictly to the eigenvalue instability of linear equations. The critical condition for flow instability as determined by linear stability analysis differs largely from the experimental data for all the three different types of flows, as shown in Table 1.

| Flow type | Re expression | Eigenvalue analysis, $\text{Re}_c$ | Experiments, $\text{Re}_c$ | Kmax at $\text{Re}_c$ (from experiments), $\equiv$ Kc |
|---|---|---|---|---|
| Pipe Poiseuille | $\text{Re} = \rho U D / \mu$ | Stable for all Re | 2000 | 385 |
| Plane Poiseuille | $\text{Re} = \rho U L / \mu$ | 7696 | 1350 | 389 |
|  | $\text{Re} = \rho u_0 h / \mu$ | 5772 | 1012 | 389 |
| Plane Couette | $\text{Re} = \rho U h / \mu$ | Stable for all Re | 370 | 370 |

Table 1 Comparison of the critical Reynolds number and the energy gradient parameter Kmax for plane Poiseuille flow and pipe Poiseuille flow as well as for plane Couette flow [17]. $U$ is the averaged velocity, $u_0$ the velocity at the mid-plane of the channel, $D$ the diameter of the pipe, $h$ the half-width of the channel for plane Poiseuille flow and plane Couette flow (Fig.1).

The proposed principle can be used for pressure and shear driven flows. If we assume that there is no energy input (such as shear) to the system or energy output from the system, this criterion can be employed to predict that the viscous flow with an inflectional velocity profile is unstable [18]. This can be explained as follow. If there is an inflection point in the velocity profile, the energy gradient in the streamwise direction from viscous friction is zero at the inflection point, while the energy gradient in the transverse direction is generally not zero. Thus, the value of K at the inflection point will be infinite, see Eq.(1). Therefore, even a small disturbance will be amplified at the inflection point by the transverse energy gradient. Following this principle, it is demonstrated that viscous parallel flow with inflectional velocity profile is a sufficient condition for flow instability for both two-dimensional and axisymmetrical flows if they are subjected to a disturbance [18]. The inviscid instability and viscous instability are compared in Table 2 [18]. Rayleigh (1880) theorem for inflection instability of inviscid was obtained only from mathematics, the physical mechanism for this theorem is still not clear.

| Theory | Flow type | Velocity profile | Flow status |
|---|---|---|---|
| Rayleigh (1880) | inviscid | no inflection | Always stable |
| | inviscid | Inflection | Always unstable |
| Energy gradient theory | viscous | no inflection | $K_{max} \leq K_c$, stable |
| | viscous | no inflection | $K_{max} > K_c$, unstable |
| | viscous | Inflection | $K_{max} = \infty$, unstable |

Table 2 Criteria of stability of parallel flows for both two-dimensional and axisymmetrical flows [18]. The critical value of K ($K_c$) depends on the flow geometry and fluid property. For both 2D and pipe Poiseuille flows, $K_c$=385[17]. For Plane Couette flow, $K_c$=370. Two-dimensional and axisymmetrical (along convex and concave surfaces) viscous flows with inflectional velocity profile are all unstable [18].

In experiments, flow transition is usually dependent on two broad factors: main flow and disturbance [1,2]. In the energy gradient theory, the condition of main flow is determined by the parameter K. This parameter can be considered as a local Re number. The condition of disturbance is determined by the average energy of disturbance because flow instability and transition is a process of energy evolution and the transition must be excited by a sufficient level of energy disturbance. We can therefore suggest that the turbulence transition depends on the energy gradient parameter of base flow and the disturbance energy.

Although laminar-to-turbulent transition can occur via several mechanisms, such as linear instability, bypass transition (skip linear instability), Gortler instability (flow on concave surface), and cross-flow instability (flow over swept wing) [19], all of these instabilities can be included or analyzed under the framework of instabilities associated with the energy gradient [17]. The total energy $E$ in Eq.(1) includes kinetic, pressure and gravitational energies for incompressible flow. A similar equation can be written for compressible flow (thermal energy and kinetic energy). A specified form of the instability may be induced by only one of the energies. For example, the instability in parallel shear flows is generated by the gradient of kinetic energy; Rayleigh-Taylor instability (a higher density fluid over a lower density fluid) is produced by the gravitational energy gradient; the cross-flow instability (flow over swept wing) is resulted from the lateral pressure gradient of the main flow (this pressure gradient is perpendicular to the streamwise direction); Gortler instability is induced by the pressure gradient across the boundary layer on a concave surface. In some occasions instability may be dominated by the mixed roles of various energies (for example, Taylor-Couette instability of flow between two rotating cylinders at high Taylor's number which is dominated by pressure and kinetic energies). In the complex process of the instability generation, some of these may be mutually enhanced or compensated. If there is no energy gradient in the flow field (uniform flow), none of the instabilities will be generated. All of the instabilities in nature are produced by the non-uniformity of the energy distribution in space and time. The evolution of natural environments can be also viewed as a result of the energy gradient roles.

## 3. Parameter K for pipe Poiseuille flow

The derivation of energy gradient parameter K for Poiseuille flows has been given in detail in [17]. Here, we just briefly describe the equations for the pipe flow for use in later sections. For pipe Poiseuille flow, the momentum equation of steady flow is written as,

$$0 = -\rho \frac{\partial p}{\partial z} + \mu \left( \frac{\partial^2 u_z}{\partial r^2} + \frac{1}{r} \frac{\partial u_z}{\partial r} \right). \tag{2}$$

This equation shows that viscous force term is proportional to the streamwise pressure gradient. The axial velocity is expressed below by integrating on the above equation,

$$u_z = u_0 \left( 1 - \frac{r^2}{R^2} \right), \tag{3}$$

where $u_0 = -\frac{1}{4\mu} R^2 \frac{\partial p}{\partial z}$ is the centerline velocity, z is in axial direction and r is in radial direction of the cylindrical coordinates, and R is the radius of the pipe. The energy gradient in the radial direction can be expressed for any position in the flow field as (noticing $u_r = 0$)

$$\frac{\partial}{\partial r} \left( \frac{1}{2} \rho V^2 \right) = -\frac{\rho}{8\mu^2} R^2 r \left( 1 - \frac{r^2}{R^2} \right) \left( \frac{\partial p}{\partial z} \right)^2, \tag{4}$$

which shows that the kinetic energy gradient in the radial direction increases quadratically with the pressure gradient.

According to the energy gradient theory, the flow instability in the pipe depends on the competition between the kinetic energy gradient in the transverse direction (Eq.(4)) and the viscous term (Eq.(2)). With increasing axial pressure gradient $\partial p / \partial z$ (or Re), the former becomes larger at a rate faster than the latter (quadratic versus linear). Therefore, at large pressure gradient (or Reynolds number), the role of energy gradient dominates over the viscous friction. This can lead to the transition to turbulence which occurs at large Re number. The energy gradient parameter K, the ratio of the two terms, clearly describes the process as Re increases.

For pipe Poiseuille flows, the ratio of the transverse energy gradient to the viscous force term, K, is ($\partial p/\partial r=0$),

$$K = \frac{\partial}{\partial r}\left(\frac{1}{2}\rho V^2\right) / \mu\left(\frac{\partial^2 u_z}{\partial r^2} + \frac{1}{r}\frac{\partial u_z}{\partial r}\right) = -\frac{2\rho r u_0}{R^2}u_0\left(1-\frac{r^2}{R^2}\right) / \left(-\mu\frac{4u_0}{R^2}\right)$$
$$= \frac{\rho U R}{\mu}\frac{r}{R}\left(1-\frac{r^2}{R^2}\right) = \frac{1}{2}\text{Re}\frac{r}{R}\left(1-\frac{r^2}{R^2}\right) \quad (5)$$

Here, $\text{Re} \equiv \frac{\rho U R}{\mu}$, $u_0$ is the maximum velocity at centerline, U is the averaged velocity and $U = \frac{1}{2}u_0$ has been used in above equation. It can be seen that K is a cubic function of radius and is proportional to Re for a fixed point in the flow field.

The distribution of u, E, and K along the transverse direction for pipe Poiseuille flow is shown in Fig.2. It is clear that there is a maximum value of K at r/R=0.5774, as shown in Fig.2. This maximum can also be obtained by differentiating Eq.(5) with r/R and setting the derivatives equal to zero. Experiments have shown that the minimum Re number for the transition to turbulence occurs at about 2000 [3,4]. Employing this experimental data in Eq.(5), we obtain $K_{max}$ =385 at r/R=0.5774. This value has been listed in Table 1.

The instability in Poiseuille flows with increasing mean velocity U is described below for a given fluid and flow geometry using the energy gradient concept. It has been suggested that the flow breakdown of the Poiseuille flow should not suddenly occur in the entire flow field, but it first takes place at the location of $K_{max}$ in the domain and then spreads out according to the distribution of K value [17]. The formation of turbulence spot in shear flows may well be explained using this procedure. In the shear flow, when the K value at a position reaches a critical value, the influence of kinetic energy gradient dominates over that of viscous friction. The flow at this position will amplify any disturbance, then transit to turbulence if the threshold disturbance is satisfied. Thus, a turbulent spot can be formed around this position. In another word, turbulence transition is a local phenomenon in the earlier stage. When the Re is further increased, this spot will grow in dimension. If the Reynolds number is sufficiently large, the full flow will transit to turbulence.

In a recent study for pipe flow, Wedin and Kerswell [20] showed that there is the presence of the "shoulder" in the velocity profile at about r/R=0.6 from their solution of the traveling waves. They suggested that this corresponds to where the fast streaks reach from the wall. It can be construed that this kind of velocity profile as obtained by simulation is similar to that of Nishioka et al's experiments for channel flows [21]. The location of the "shoulder" is about same as that for $K_{max}$. According to the present theory, this "shoulder" may then be intricately related to the distribution of energy gradient.

For a plane Poiseuille flow, the maximum value of K is located at y/h=±0.5774, fairly similar to pipe flow. This position should then be the most dangerous location for flow breakdown, which has been confirmed by Nishioka et al's experiment [21]. Nishioka et al's [21] experiments for plane Poiseuille flow showed details of the outline and process of the flow breakdown. The measured instantaneous velocity distributions indicate that the first oscillation of the velocity occurs at y/h=0.50~0.62.

As can be noted from Eq.(5), K is proportional to the global Reynolds number. A large value of K has the ability to amplify the disturbance, and vice versa. Therefore, at high Re, the disturbance energy needed to trigger transition becomes correspondingly smaller.

## 4. Amplitude of disturbance for transition in a pipe flow

### 4.1 Disturbance energy and the energy angle at critical condition

Dou [17] proposed the concept of energy angle based on the theory of energy gradient. The relation of parameter K with the energy angle $\alpha$ follows from Eq.(1a), (see also Fig.1):

$$K = \frac{(\nabla E)_r}{(\nabla E)_z} = \tan\alpha = \frac{1}{\tan\beta} \tag{6}$$

or

$$\tan\beta = \frac{1}{K}, \tag{7}$$

where $\alpha$ is the so called "energy angle," and $\beta$ is the complementary angle of $\alpha$. It has been suggested that,

$$\begin{matrix} \alpha < \alpha_c, & (\beta = 90° - \alpha), & \textbf{the flow is stable;} \\ \alpha \geq \alpha_c, & (\beta = 90° - \alpha), & \textbf{the flow is unstable;} \\ \alpha = 90° & (\beta = 0°), & \textbf{the flow is unstable.} \end{matrix} \tag{8}$$

where,

$$\alpha_c = \arctan K_c. \tag{9}$$

Here $\alpha_c$ is called the critical energy angle for flow transition. When $K \to \infty$, $\alpha \to 90°$ and $\beta \to 0°$. At this condition, the disturbance energy needed to trigger the transition is infinitely small, i.e, $T \to 0$. Here, T expresses the threshold disturbance energy needed to trigger the transition at a given base flow (and thus Re is given).

Now, we use the subscript *t* to express the critical condition at which the disturbance energy needed to trigger the transition is infinitely small; then the corresponding parameters are $\text{Re}_t$, $K_t$, $\beta_t$, and $T_t$ (actually $T_t = 0$), see also Fig.3. For the pipe Poiseuille flow, we know that $\text{Re}_t = \infty$ from the linear analysis [1,2]. Thus, we have $K_t = \infty$ and $\beta_t = 0$ from Eqs.(5) and (6). For the plane Poiseuille flow, we know that $\text{Re}_t = 5772$ again from the linear analysis [1,2], and thus $K_t \neq \infty$ and $\beta_t \neq 0$ from Eqs.(5) and (6).

As discussed earlier, **the turbulence transition depends on the energy gradient parameter of base flow and the disturbance energy**. Since the energy gradient parameter of base flow can be expressed by the angle $\beta$ (Eq.(6)), the threshold disturbance energy T needed to trigger the transition can therefore be expressed as a function of angle $\beta$. Thus, T is expressed as

$$T = f(\beta). \tag{10}$$

Next, we expand T as a Taylor series at the vicinity of $\beta_t$,

$$T = f(\beta_t) + a_1(\beta - \beta_t) + a_2(\beta - \beta_t)^2 + a_3(\beta - \beta_t)^3 + \cdots\cdots, \quad (\beta - \beta_t \ll 1), \quad (11)$$

where, $a_n = f^{(n)}(\beta_t)/(n!)$ and $f^{(n)}(\beta_t)$ is the derivatives of nth order at $\beta_t$.

Since $T_t = f(\beta_t) = 0$ as discussed above for the pipe Poiseuille flow, Eq.(11) is rewritten below with $\Delta\beta = \beta - \beta_t$ as,

$$T = a_1\Delta\beta + a_2\Delta\beta^2 + a_3\Delta\beta^3 + \cdots\cdots, \tag{12}$$

where $a_1, a_2, \ldots a_n \ldots$ are dimensional constants.

From the experiments in pipe flow, it is known that the minimum Re number for the turbulent transition is about 2000, and $K_{max}$ is 385 as shown in Table 1. At this condition, $\Delta\beta = 0.15°$ from Eq.(6). For $\text{Re} > 2000$, $\Delta\beta$ becomes smaller than $0.15°$ as it decreases with increasing Re.

Since $\Delta\beta \ll 1$ generally, we can just utilize the linear part in Eq.(12) at least in the first order approximation,

$$T = a_1\Delta\beta. \tag{13}$$

Furthermore, we have

$$\Delta\beta \approx \tan\Delta\beta = \tan(\beta - \beta_t). \tag{14}$$

Because $\tan\beta \tan\beta_t \ll 1$, we obtain

$$\tan\Delta\beta = \frac{\tan\beta - \tan\beta_t}{1 + \tan\beta \tan\beta_t} \approx \tan\beta - \tan\beta_t. \tag{15}$$

Therefore, with the relation that $\tan\beta = 1/K$ and $\tan\beta_t = 1/K_t$ (Eq.(6)), the following equation is obtained from Eqs. (13), (14) and (15),

$$T = a_1\Delta\beta = a_1\tan\Delta\beta = a_1(\tan\beta - \tan\beta_t) = a_1\left(\frac{1}{K} - \frac{1}{K_t}\right). \tag{16}$$

Since the parameter $K$ is proportional to the global Reynolds number for a given geometry (see Eq.(5)), we have $K \propto \text{Re}$ and hence

$$T = b_1\left(\frac{1}{\text{Re}} - \frac{1}{\text{Re}_t}\right), \tag{17}$$

where $b_1$ is another constant.

## 4.2 Calculation of average energy of disturbance in a period
### 4.2.1 Average energy of sine wave disturbance function

In the (usual) experiments on turbulence transition, the initial disturbance to the pipe flow is generally produced at the inlet section by various means. In later sections, we will compare the theory with the experiments in [6]. In order to better understand the derivation below, a brief introduction to the experiment in [6] is given. In the experiment, the disturbance is introduced into the pipe by means of jet(s) which is driven from a motor and piston system. A valve is mounted at the outlet of the device to stops fluid being sucked out of the pipe system on the return stroke of the piston. There are two kinds of disturbance inlet geometries used in the experiments. The first is that the disturbance at the pipe is in the form of a single jet injected orthogonally to the main stream flow through a small hole opening on the pipe wall. The second geometry consists of an arrangement of valves and tubes which are used to produce multiple (six) jets of small diameter directed azimuthally to the pipe flow so as to introduce a component of swirl into the main flow field. Two parameters A (amplitude) and $\omega$ (disturbance frequency) along with the diameters of the inlets to the pipe characterize the disturbances in terms of the average flow rate of the disturbance and total mass flux added to the main flow. The parameter A is the maximum amplitude of the disturbance taken to be the distance of the travel of the piston. The speed of disturbance jet is almost linear after the initial impulse.

For comparison, we first give the calculation of disturbance energy for a disturbance associated with a sinusoidal movement of the piston. That is, the motion of the piston is expressed by the equation,

$$Z = A\sin(\omega t + \varphi_0), \tag{18}$$

where A is the maximum axial movement of the piston (which has exactly the same meaning as that in the experiments of Darbyshire and Mullin [6]), $\omega$ is the frequency of the disturbance, $t$ is the time, and $\varphi_0$ is the phase angle at initial position. Thus, the velocity of the piston can be expressed as a cosine function (Fig.4),

$$\dot{Z} = A\omega\cos(\omega t + \varphi_0). \tag{19}$$

The added fluid volume to the pipe by injection should be equal to the fluid volume displaced out by the piston. The maximum disturbance in volume due to the piston motion is $AS_1$, where $S_1$ is the cross-sectional area of the piston. The corresponding maximum disturbance in volume in the pipe is the same as that associated with the piston, $AS_1 = BS_2$. Here B is the maximum disturbance distance in the pipe under the maximum volume disturbance, and $S_2$ is the cross-sectional area of the pipe. From this relation, we have the following result,

$$B = AS_1 / S_2. \tag{20}$$

Thus, corresponding to Eq.(18), the axial disturbance in the pipe is

$$x = B\sin(\omega t + \varphi_0). \tag{21}$$

The velocity disturbance in the pipe is obtained as,

$$\dot{x} = B\omega\cos(\omega t + \varphi_0) \tag{22}$$

or

$$\dot{x} = u_A \cos(\omega t + \varphi_0), \; u_A = B\omega. \tag{23}$$

Thus, the average kinetic energy for unit mass fluid in a period is,

$$T = \frac{1}{\tau}\int_0^\tau \frac{1}{2}\dot{x}^2 dt = \frac{1}{2}\frac{B^2\omega^2}{\tau}\int_0^\tau \cos^2(\omega t + \varphi_0)dt = \frac{1}{4}B^2\omega^2 \tag{24}$$

where $\tau = 2\pi/\omega$ is the period of the disturbance.

### 4.2.2 Average energy of disturbance in Darbyshire and Mullin's pipe flow experiments

In the experiments of Darbyshire and Mullin[6], the disturbance input is done with jet injection at the inlet part of the pipe which is associated with the piston movement. The piston is driven by a mechanical mechanism with a periodic motion. The velocity disturbance due to the movement of the piston is not a cosine function. The velocity variation of the piston with time can be expressed as (pulse-linear disturbance, see Fig.4),

$$\left.\begin{array}{ll} \dot{z} = A\omega\left(1 - \dfrac{\omega t}{\pi}\right) & 0 < \omega t < \pi \\ \dot{z} = 0 & \pi < \omega t < 2\pi \end{array}\right\}. \tag{25}$$

Thus, the disturbance velocity in the pipe is

$$\left.\begin{array}{ll} \dot{x} = B\omega\left(1 - \dfrac{\omega t}{\pi}\right) & 0 < \omega t < \pi \\ \dot{x} = 0 & \pi < \omega t < 2\pi \end{array}\right\}. \tag{26}$$

The disturbance energy in the pipe can be calculated as

$$T = \frac{\omega}{2\pi}\left[\int_0^{\pi/\omega}(B\omega)^2\left(1-\frac{\omega t}{\pi}\right)^2 dt + \int_{\pi/\omega}^{2\pi/\omega} 0 dt\right] = \frac{\omega}{2\pi}(B\omega)^2\int_0^{\pi/\omega}\left(1-\frac{\omega t}{\pi}\right)^2 dt = \frac{1}{6}B^2\omega^2. \tag{27}$$

It is seen from Eqs.(24) and (27) that the average kinetic energy for unit mass fluid in a period for pulse-linear disturbance in [6] is less than that for cosine velocity disturbance, and the latter is 1.5 times of the former.

### 4.3 Threshold amplitude versus the Re

Introducing Eq.(27) into Eq.(17), we obtain

$$\frac{1}{6}B^2\omega^2 = b_1\left(\frac{1}{\text{Re}} - \frac{1}{\text{Re}_t}\right). \tag{28}$$

Substituting Eq.(20) into Eq.(28), we have

$$\frac{1}{6}\left(A\frac{S_1}{S_2}\right)^2\omega^2 = b_1\left(\frac{1}{\text{Re}} - \frac{1}{\text{Re}_t}\right). \tag{29}$$

For a given $S_1$ (the area of the cross-section of the piston) and $S_2$ (the area of the cross-section of the pipe), we obtain

$$A^2\omega^2 \propto \left(\frac{1}{\text{Re}} - \frac{1}{\text{Re}_t}\right), \tag{30}$$

and furthermore,

$$A^2 \propto \frac{1}{\omega^2}\left(\frac{1}{\text{Re}} - \frac{1}{\text{Re}_t}\right). \tag{31}$$

For a given disturbance frequency $\omega$, therefore we have

$$A^2 \propto \left(\frac{1}{\text{Re}} - \frac{1}{\text{Re}_t}\right). \tag{32}$$

In the above equation, $\text{Re}_t$ is the Reynolds number at which the flow transits to turbulence by infinitesmall disturbance. Because $\text{Re}_t = \infty$ for a pipe Poiseuille flow at infinitesmall disturbance as discussed earlier, we then obtain from Eq.(32) as

$$A \propto \frac{1}{\text{Re}^{1/2}}. \tag{33}$$

It should be restated that the parameter A is the same as that employed in Darbyshire and Mullin [6], i.e., the disturbance amplitude of the piston. If the disturbance amplitude is normalized by the average velocity in the pipe, the above expression can be written as

$$\overline{A} = \frac{A}{U} \propto \frac{1}{\text{Re}^{3/2}}. \tag{34}$$

Substituting Eq.(20) into Eq.(34), the disturbance amplitude in the pipe normalized by the average velocity becomes

$$\overline{B} = \frac{B}{U} \propto \frac{1}{\text{Re}^{3/2}}. \tag{35}$$

This result of Eq.(34) or Eq.(35) is in agreement with Chapman [13], Meseguer [15], and Trefethen et al [14]; all attained a component of $\gamma = -3/2$ for the pipe Poiseuille flow.

## 5. Further comparison of theory with experiments

Equation (33) is depicted in Fig.5 and compared with Darbyshire and Mullin' [6] experiments. The magnitude of the proportional constant in Eq.(33) does not feature for understanding the mechanism as stated by Meseguer [15]. It can be found that the disturbance amplitude decreases with Re for both the theory and experiments. Equation (33) has shown good agreement with the experimental data for the range of large Re (Re=2700--10000). This provides support that the present model is reasonable in the large Re range. However, the trend of amplitude variation versus Re has still not been satisfactorily explained thus far. Some numerical simulations with traveling waves in pipe flow have indicated that the development of these traveling waves ultimately harbors a turbulent attractor as Re increases [22,23].

According to the energy gradient theory [17], the mechanism of disturbance vis-a-vis Reynolds number (Fig.5) can be explained as follows. With increasing Re, K increases and the energy angle $\alpha$ also increases. Thus, a large transverse energy gradient facilitates the energy exchange between the fluid layers. This exchange of energy will amplify the disturbance in the flow and finally lead to transition if both the energy gradient and the disturbance amplitude are sufficiently large. At higher energy angle condition, the ability of mean flow to amplify the disturbance becomes greater. Thus, the initial disturbance energy needed for the transition is correspondingly reduced.

From Fig.5, the experimental data show that the exponent $\gamma$ varies at different Re range, and it can not be represented by a single constant in the Re range considered. Therefore, it is commented that the analysis and the exponent found in Chapman [13] may not be universal. Meseguer's [15] simulations (Fig.8b) also show that the critical amplitude can not be expressed by a single exponent in the whole Re range, although some recent experiments with a pulse disturbance have indicated that $\gamma = -1$ for a large Re range [16]. Further studies are needed to clarify these problems.

Darbyshare and Mulin's experiments [6] showed that for a given Re (so is $K_{\max}$), the critical amplitude also depends on the frequency of the disturbance. The experiments with a single jet injection showed that the critical amplitude *A* decreases with increasing frequency. This suggests that the disturbance energy is a key factor. The experimental data also showed that this critical energy of disturbance is not a constant with the variation of frequency at a fixed Re. The experimental results at Re=2200 in [6] are depicted in Fig.6. The symbols are the experimental data for critical amplitude *A* versus the frequency of the injection. A best correlation is achieved with $\omega^2 A$=constant as indicated by the solid line drawn in the same figure. Thus, the disturbance energy is $T = \omega^2 A^2 = CA$ with C as a constant, which shows that the critical energy is proportional to the amplitude *for a given Re* when the disturbance frequency is allowed to vary. The mechanism or mechanics for this behaviour is not clear, and may be related to the complex process of turbulence generation, which needs further investigation. In the above relation, it should be noted that *A* is not infinitely small, otherwise the disturbance frequency will be infinite.

Although a few studies have been carried out on the mechanism of transition, the detailed development and process during the transition have still largely not been clarified or fully understood. In fact, the transition to turbulence is considered by some to be a redistribution of the energy in the flow. As discussed earlier, the origin and the process in the transition are all dominated by the energy gradient. The excitation of the process is initiated and attributed to the disturbance. A sufficient magnitude of energy in the disturbance is a necessary condition for transition. Since the disturbance energy is proportional to the amplitude and the square of the frequency, the critical condition is also related to the frequency. For $Re$ larger than the minimum transition Reynolds number, $Re_c$, a small disturbance may not be enough to stimulate the transition, but this disturbance can lead to oscillation of the base flow (incompressible). This flow oscillation may be associated with the Tollmien-Schlichting waves with resulting streamwise streaks. For the said Reynolds number range, if the disturbance energy is sufficiently large (i.e., exceeding the threshold value), the flow may "skip" the Tollmien-Schlichting waves (namely bypass transition). At this large disturbance, the base flow may be lead to a velocity profile inflectional such that the flow transits to turbulence via three-dimensional vortex diffusion mechanism. Following the idea developed in this paper, the route of the events during the transition is suggested for the different initial disturbances:

**Transverse energy gradient +initial disturbance→ amplification of disturbance→ (1) or (2):**
**(1) (small initial disturbance) →Tollmien-Schlichting waves→streamwise streaks**
**(2) (large initial disturbance) →Inflection velocity profile→ 3D vortex formation→bypass transition**

## 6. Concluding Summary

Energy gradient theory was proposed for laminar-turbulence instability [17]. It has been demonstrated that this theory is valid at least for parallel flows in [17]. In this paper, based on the energy gradient theory, we argued that the transition to turbulence depends on the energy gradient of mean flow and the disturbance energy. Furthermore, the threshold disturbance energy required for turbulence transition is expressed as a function of the complementary angle ($\beta$) to the energy angle ($\alpha$) which characterizes the behaviour of the energy gradient of mean flow. Employing the Taylor series expansion of the functional and some simplifications, the variation of the threshold amplitude of disturbance versus the Reynolds number is derived. In the derivation, the effect of frequency of disturbance on the transition is shown and which is depicted by the disturbance energy being proportional to the square of the disturbance frequency. It is found that the normalized amplitude of disturbance scales with Re by an exponent of -3/2 for the transition condition in pipe flow. This finding is in agreement with Chapman [13], Meseguer [15], and Trefethen et al [14]. From the theory and the coupled with experimental data analysis, it is observed that the energy of the disturbance is the key factor to excite the transition rather than only the amplitude. Finally, it is suggested that the energy gradient theory presents yet another possible approach for the description of the turbulent transition. In addition, the phenomenon of variation of the disturbance amplitude A versus the disturbance frequency $\omega$ for a given Re, $\omega^2 A$ =constant, which is deduced from experimental data, needs further investigation.

**Acknowledgement**

The authors are grateful to Prof. LN. Trefethen (Oxford University) for his helpful comments on the first version of this paper.

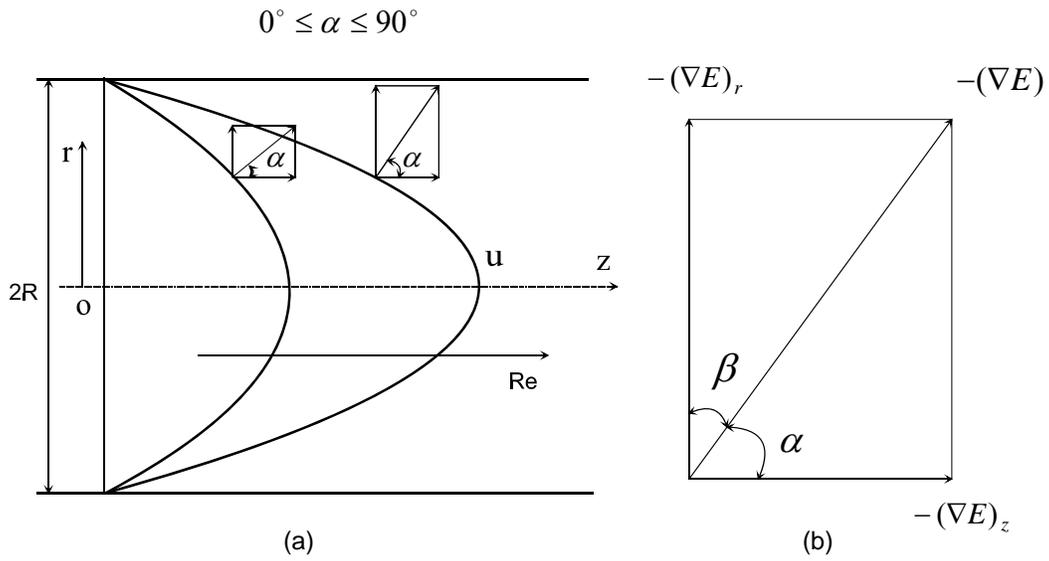

Fig.1 (a) Velocity distribution with the Re in the pipe. (b) Definition of energy angle

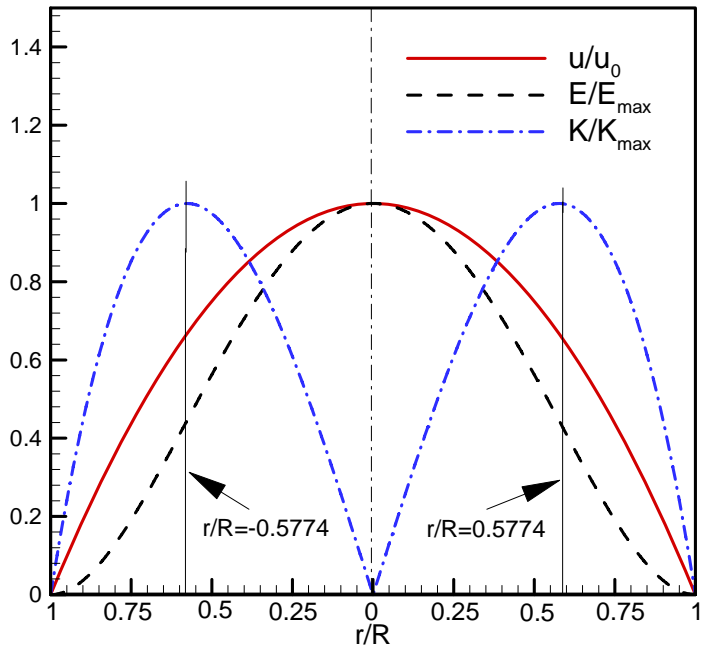

Fig. 2 Velocity, energy, and K along the transverse (radial) direction for pipe Poiseuille flow, which are normalized by the their respective maximum.

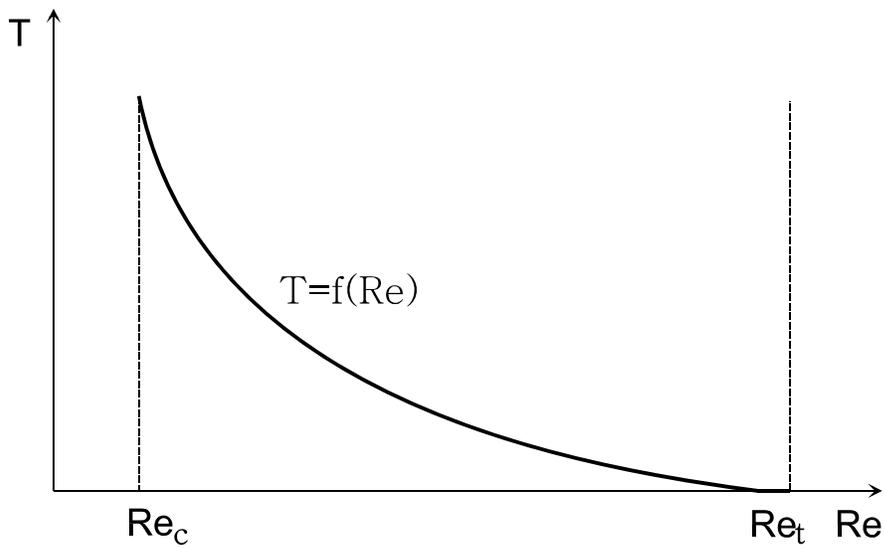

Fig.3 Threshold of disturbance energy for transition to turbulence in a pipe versus the Reynolds number. $Re_c$ is the minimum Reynolds number at which the flow transits to turbulence. $Re_t$ is the Reynolds number at which the flow transits to turbulence by infinite small disturbance. For pipe Poiseuille flow, $Re_c = 2000$ and $Re_t = \infty$. For plane Poiseuille flow, $Re_c = 1000$ and $Re_t = 5772$.

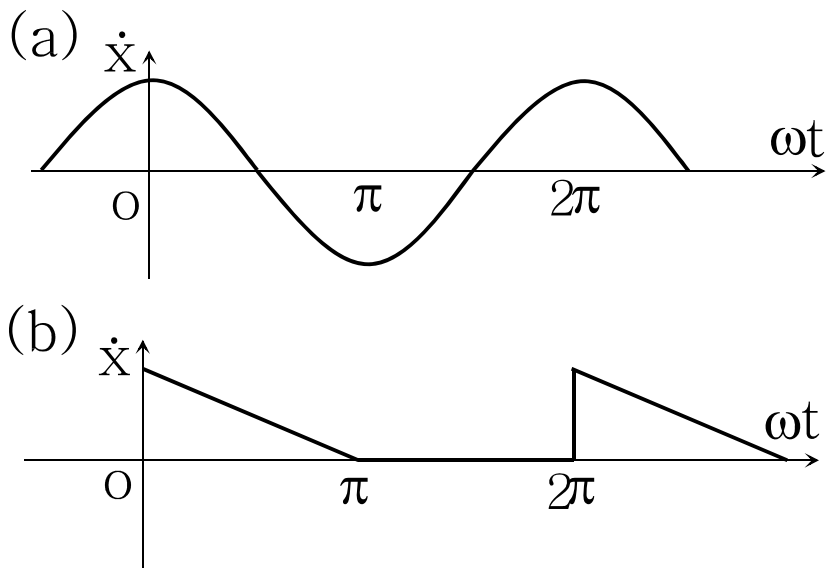

Fig.4 (a) Cosine function for the velocity disturbance. (b) Disturbance function in Darbyshire and Mullin [6]'s experiments.

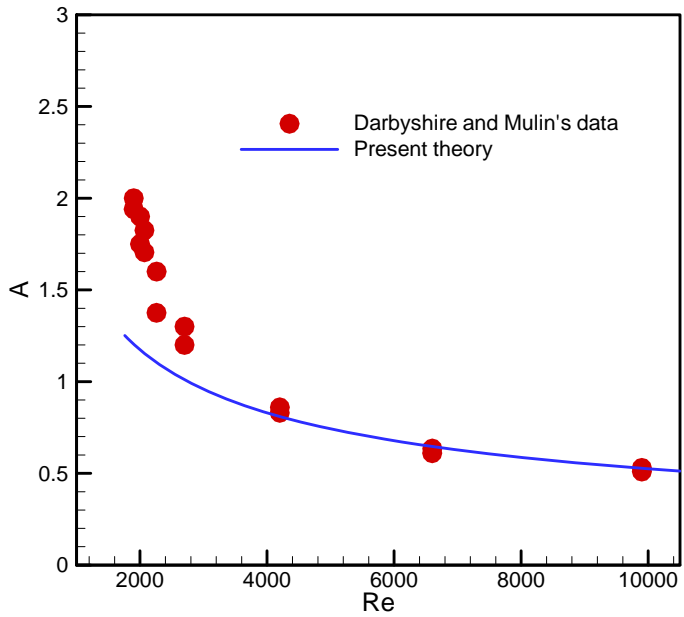

(a)

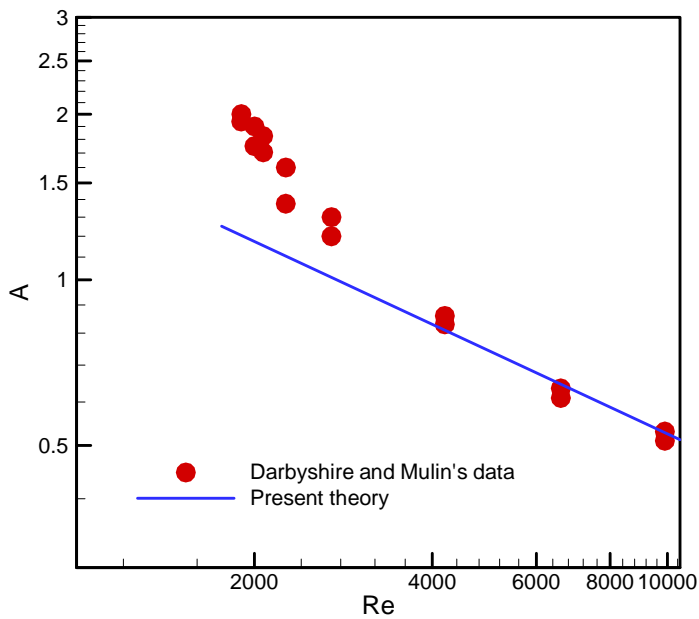

(b)

Fig.5 The amplitude of disturbance versus the Re number $A \propto \mathrm{Re}^{-1/2}$ (Eq. (33)). Fig.5(a) and Fig.5(b) are presented with different ordinate scaling. For each Re, the two solid circle symbols express the limits of experiments uncertainty of the critical value, which are taken from Figs.14 and 18 of Darbyshire and Mullin [6] for the experiments using the six-jet disturbance at constant drive speed (frequency) of $\omega = 0.409\,s^{-1}$.

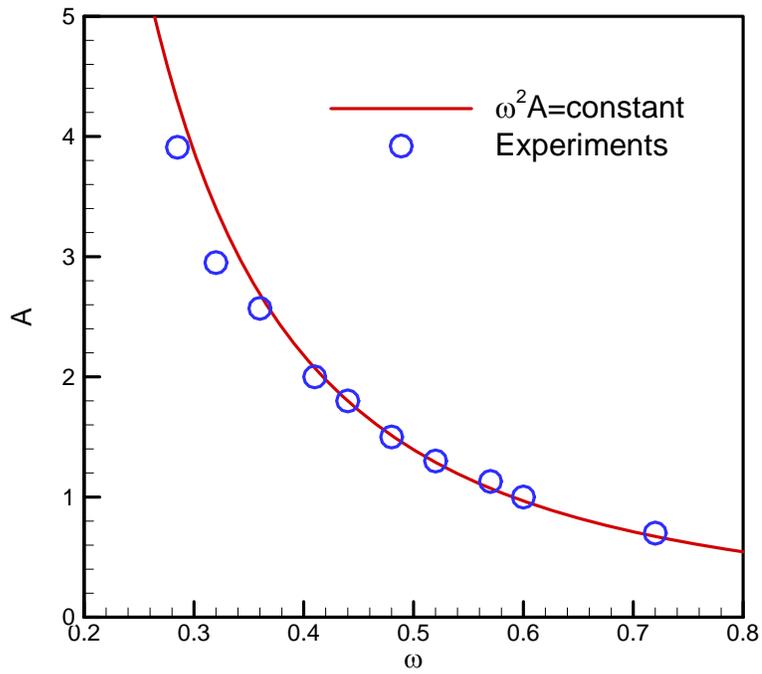

Fig. 6 Variation of critical disturbance amplitude, $A$, against the driving frequency $\omega$ at Re=2200 for the single-jet disturbance. Correlation for the experimental data [6] is given as depicted by the solid line.